\documentclass[10pt]{article}

\author{Keiichi Shigechi\footnote{E-mail address: shigechi@monet.phys.s.u-tokyo.ac.jp} 
and Miki Wadati\footnote{E-mail address: wadati@monet.phys.s.u-tokyo.ac.jp} \vspace*{0.2cm}\\
\textit{Department of Physics, Graduate School of Science} \\
\it{University of Tokyo, Hongo 7-3-1, Bunkyo-ku, Tokyo, 113-0022, Japan}\vspace*{0.5cm}\\
Ning Wang\footnote{E-mail address: wangyu@public.qd.sd.cn}\vspace*{0.2cm}\\
\it{Department of Physics, Ocean University of China}\\
\it{5-Yushan Rd. Qingdao, 266003 P.~R.~China}\vspace*{0.2cm}\\}
\title{WDVV equation and Triple-product Relation}
\date{22 April, 2004}

\begin{document}

\maketitle

\begin{abstract}
We study the relation between the WDVV equations and the $\tau$-function of the noncommutative KP (NCKP) hierarchy. 
WDVV-like equations (Hirota triple-product relation) in the noncommutative context appear as a consequence of the 
non-trivial equation for $\tau$-function of the NC KP hierarchy, while
the prepotential in the Seiberg-Witten (SW) theory has been identified to the $\tau$-function of the Whitham hierarchy.
We show that the spectral curve for the SW theory is the same as the Toda-chain hierarchy. We also show that Whitham hierarchy 
includes commutative Toda/KP hierarchy as a construction. Further, we comment on the origin of the Hirota triple-product 
relation in the context of 
the SW theory.
\end{abstract}

\section{Introduction}
The low energy effective action for the super Yang-Mills theory was introduced by Seiberg and Witten~\cite{SW}(see also\cite{Ler}).
There, the data of the effective action are inverted to that of the prepotential and the Riemann surfaces. Subsequently,  
after, the correspondence between the supersymmetric Yang-Mills theory and the integrable systems, such 
as Hitchin systems~\cite{Don1}, Toda systems~\cite{Mir6,War1}, Calogero-Moser systems~\cite{Mor1,DH} was pointed out. In this context, 
the spectral curves which we construct from the Lax representation of the integrable systems are the same hyperelliptic curves 
(not all hyperelliptic in some gauge groups) as the case of the Seiberg-Witten consideration. On the other hand, the prepotential 
is extended to the one with ``Whitham times,'' that is the logarithm of the $\tau$-function of the Whitham hierarchy. 
The Whitham hierarchy gives us important perspectives in terms of the RG equations~\cite{Mir4} and the instanton correction for 
the prepotential~\cite{Mari1}. 


The Witten-Dijkgraaf-Verlinde-Verlinde (WDVV) equation was first studied in the two dimensional gravity~\cite{WDVV1,WDVV2}. 
This equation arises as a consequence of the crossing relations which are satisfied by the structure constants of the operator 
algebra of primary operators in Landau-Ginzburg model. It was shown that the third derivatives of the prepotential obey the WDVV 
equations. 

It was observed by Krichever~\cite{Kri1} that for curves of genus zero the $\tau$-function of a `` universal" Whitham hierarchy gives a 
solution to the WDVV equations. We now know that any $\tau$-function of the dispersionless KP/Toda hierarchies provide a solution to WDVV equations in the context of Whitham hierarchies~\cite{Mar2}.


The Weyl-Moyal product is a powerful tool to study the noncommutative field theory. However, it is difficult to construct the 
noncommutative $\tau$-function through the Weyl-Moyal context~\cite{Paniak:2001fc}. The other approach for the noncommutative 
theory is the operator formalism. Wang and Wadati~\cite{WW} studied $\tau$-function representation of solutions and 
deformation of Hirota bilinear relations for a noncommutative Kadomtsev-Petviashvili (NC KP) hierarchy. The $\tau$-function 
for the NC KP hierarchy satisfies an additional triple-product relation (related to WDVV equation, see sect.5).  


Our goal in this paper is to reveal the relation between WDVV equation in the context Seiberg-Witten/Whitham theory and the non-trivial 
triple-product relation for a NC KP hierarchy.

The paper consists of the following. In sect.2, we briefly remind the Seiberg-Witten theory for the pure gauge group and the prepotential of the SW theory. For simplicity, we focus on the $SU(N)$ gauge group without matter.  
In sect.3, we also briefly derive the WDVV equation following Marshakov et.al~\cite{Mir1,Marshakov:1996ae}.
In sect.4 The prepotential in the S-W theory ($SU(N)$ case) is discussed by use of the Whitham hierarchy. 
See~\cite{Taka2} for the other gauge groups. We construct the Whitham hierarchy via 
the data on the Riemann surface. Integrable systems such as KdV and Toda/KP hierarchy are also considered in terms of the Whitham hierarchy.
In sect.5, NC KP equation is discussed through the operator-valued $\bar{\partial}$-method. Almost all the formulae of the 
noncommutative $\bar{\partial}$-problem remain to be the same as those of commutative case. We remark, however, that 
the non-trivial equation for the noncommutative $\tau$-function arises. The last section is devoted to concluding remarks. 

\section{Seiberg-Witten Solution of super Yang-Mills Theory}
The data of the Seiberg-Witten solution of the SU(N) pure gauge theory are given as follows~\cite{SW}. To be specific, we consider the 
$d=4\ \mathcal{N}=2$ supersymmetric Yang-Mills theory without matter multiplets. 
The Seiberg-Witten curve $\Gamma$ for the gauge group $SU(N)$ is a complex algebraic curve of the form,
\begin{equation}
w+\frac{\Lambda^{2N}}{w}=P_{N}(\lambda),
\end{equation}
or equivalently in hyperelliptic parameterization,
\begin{equation}
y^{2}=P_{N}(\lambda)^{2}-4\Lambda^{2N}, \label{cc1}
\end{equation}
where $P_{N}(\lambda)\equiv\lambda^{N}-\sum_{k=2}^{N}u_{k}\lambda^{N-k}$ is a polynomial of degree $N$, $\Lambda$ is a scale 
parameter of the theory, which can be put equal to unitary, and $y=w-\frac{\Lambda^{2N}}{w}$. The genus of this spectral curve is 
$N-1$. We denote canonically normalized cycles on the curve by $A_{i}, B_{i} (i=1,\cdots,N-1)$ with $A_{i}\circ B_{j}=\delta_{ij}$. 

Let us introduce the prepotential $\mathcal{F}(a_{i},\Lambda)$. The prepotential contains all the information about the 
low energy limit of the $\mathcal{N}=2$ SUSY YM theory. The prepotential is holomorphic and has the following properties,
\begin{equation}
a_{i}^{D}=\frac{\partial\mathcal{F}}{\partial a_{i}}
\end{equation}
where
\begin{equation}
a_{i}=\oint_{A_{i}}dS_{SW},\ \ \ a_{i}^{D}=\oint_{B_{i}}dS_{SW}.
\end{equation}
The meromorphic differential $dS_{SW}$ is defined as 
\begin{equation}
dS_{SW}=\lambda d\log w.
\end{equation}
Note that 
\begin{equation}
\frac{\partial dS_{SW}}{\partial a_{i}}=dw_{i},\ \ \ \frac{\partial^{2}\mathcal{F}}{\partial a_{i}\partial a_{j}}
=\frac{\partial a_{i}^{D}}{\partial a_{j}}=\mathcal{T}_{ij}
\end{equation}
where $dw_{i}$ are canonical holomorphic differentials and $\mathcal{T}_{ij}$ is the period matrix of the complex 
curve (\ref{cc1}).

In the theory of the integrable systems, the complex curve (\ref{cc1}) is obtained as the spectral curve for the periodic Toda system.
The coefficients of $\lambda^{i}\ (i=1,\cdots N-1)$ are the functions of the $\mathrm{tr}(L^{k})\ (k=1,\cdots N-1)$, where $L$ is a 
matrix of the Lax representation of the Toda systems. In particular, the coefficient of $\lambda^{N-1}$ in $P_{N}$ corresponds to $\mathrm{tr}L=0$.

\section{WDVV equations}

The WDVV equations are due to the associativity in the theory of 2d gravity~\cite{WDVV1,WDVV2}.
In two-dimensional topological theories such as  $\mathcal{N}=2$ SUSY Landau-Ginzburg model, the WDVV equations are the 
consequence of the crossing relations 
\begin{equation}
\sum_{k}C_{ij}^{k}C_{kl}^{n}=\sum_{k}C_{il}^{k}C_{kj}^{n}
\end{equation}
for the structure constants of the operator algebra among primary operators $\{\phi_{j}\}$ which satisfy the relation 
$\phi_{i}\cdot\phi_{j}=\sum_{k}C_{ij}^{k}\phi_{k}$. 

Here, we consider the WDVV equation as a result of commutative conditions for some matrices. We follow the derivation 
of Marshakov et.al~\cite{Mir1,Marshakov:1996ae}(see also~\cite{Mir6,Mir2,Mar1,Mir3}).
Let us first define a matrix with index $i$ as the third derivatives of the holomorphic function (or prepotential in the SW theory)
\begin{equation}
(F_{I})_{MN}\equiv \frac{\partial^{3}\mathcal{F}}{\partial a_{I}\partial a_{M}\partial a_{N}},
\end{equation}
where $\mathcal{F}=\mathcal{F}(a_{i})$ is a function of quantum moduli $a_{i}(i=1,\cdots,n)$.

We impose the commutation relation for the matrices $C_{I}^{(Q)}(I=1,\cdots,n)$.
\begin{equation}
C_{I}^{(Q)}C_{J}^{(Q)}=C_{J}^{(Q)}C_{I}^{(Q)}\ \ \ \forall I,J
\label{ComCond}
\end{equation}
where the indices $Q$ refer to the ``metric" matrix,  
\begin{equation}
(Q)_{MN}=\sum_{K}q_{K}(F_{K})_{MN}.
\end{equation}
Matrices $C_{I}^{(Q)}$ are defined via $Q$ and $F_{I}$.
\begin{equation}
C_{I}^{(Q)}\equiv Q^{-1}F_{I},\ \mathrm{or}\ \ C^{(Q)M}_{IN}=(Q^{-1})^{ML}F_{ILN}.
\label{DefC}
\end{equation}

If we take $Q=F_{K}$ and substitute it into Eq.(\ref{DefC}), we have $C_{I}^{(K)}=F_{K}^{-1}F_{I}$. Then, Eq.(\ref{ComCond}) is
converted to the standard form
\begin{equation}
F_{I}F_{K}^{-1}F_{J}=F_{J}F_{K}^{-1}F_{I}\ \ \ \forall I,J, 
\label{WDVV}
\end{equation}
which we call the WDVV equations.
\section{Whitham Hierarchy}
\subsection{Formulation}
The Whitham hierarchy is based on the following data~\cite{Kri1}. Let $\Omega_{I}(k,t_{I})$ be a 
set of holomorphic functions of the variable $k$, which is defined in some complex domain. On the space with coordinates 
$(k,t_{I})$, we introduce a 1-form
\begin{equation}
w=\sum_{I}\Omega_{I}(k,t_{I})d t_{I}.
\end{equation}
Its full external derivative equals $\displaystyle d w=\sum_{I}d\Omega_{I}(k,t_{I})\wedge dt_{I}$, where 
$d w=\partial_{k}\Omega_{I}d k\wedge d t_{I}+\partial_{J}\Omega_{I}d t_{I}\wedge d t_{J}$. One can define 
general Whitham equations (or Whitham hierarchy) as
\begin{equation}
d w\wedge d w=0 \label{WhithamH}
\end{equation}

The algebraic form Eq.(\ref{WhithamH}) of the Whitham equations is equivalent to 
\begin{equation}
\sum \partial_{k}\Omega_{[I}\partial_{J}\Omega_{C]}=0
\end{equation}
where a symbol $[\cdots]$ means antisymmetrization. For some fixed index $I=I_{0}$, we introduce the `` Darboux"
coordinates,
\begin{equation}
t_{I_{0}}=x,\ \ \ \Omega_{I_{0}}(k,t_{I})=p.
\end{equation}
One can get from Eq.(\ref{WhithamH}) the standard form,
\begin{eqnarray}
&&\partial_{I}\Omega_{J}-\partial_{J}\Omega_{I}+\{\Omega_{I},\Omega_{J}\}=0, \label{WhithamEq} \\
&&\{\Omega_{I},\Omega_{J}\}\equiv \frac{\partial\Omega_{I}}{\partial x}\frac{\partial\Omega_{J}}{\partial p}
-\frac{\partial\Omega_{J}}{\partial x}\frac{\partial\Omega_{I}}{\partial p},
\end{eqnarray}

Equation (\ref{WhithamEq}) is considered as the compatibility conditions of the system,
\begin{equation}
\frac{\partial \lambda}{\partial t_{I}}=\{\lambda,\Omega_{I} \},
\end{equation}
and equivalently
\begin{eqnarray}
\frac{\partial \Omega_{I}(\lambda,t)}{\partial t_{J}} \big|_{\lambda}
=\frac{\partial \Omega_{J}(\lambda,t)}{\partial t_{I}}\big|_{\lambda}.
\end{eqnarray}
The compatibility conditions guarantee the existence of potential $S$ such that
\begin{equation}
\Omega_{I}(\lambda,t)|_{\lambda}=\frac{\partial S(\lambda,t)}{\partial t_{I}}\big|_{\lambda},
\label{Omega}
\end{equation}
and therefore the 1-form can be written as 
\begin{equation}
w=d S-\frac{\partial S(\lambda,t)}{\partial \lambda}d\lambda.
\end{equation}

The Whitham equations describe the dynamics on moduli space of complex curves where the Whitham times correspond to the moduli.
To realize this, we choose 1-differentials $d\Omega_{I}$ to be holomorphic outside the marked point where the differentials have 
fixed behaviors. The Whitham hierarchy associated with the finite-gap solution of the KdV hierarchy is related to 
the hyperelliptic curve $y^{2}=R(\lambda)=\sum_{\beta}^{2g+1}(\lambda-r_{\beta})$~\cite{Flashka}. 

The following set of data is necessary to get non-trivial gap solutions to
Whitham hierarchy:
\begin{itemize}
\item Riemann surface (complex curve) of genus g
\item a set of marked points $P_{i}$ (punctures)
\item coordinates $k_{i}$ in the vicinities of the marked points $P_{i}$
\item a pair of differentials $(d\lambda,dz)$ 
\end{itemize}

We first consider the case with a single puncture. This case is typical, and for example, describes the Whitham hierarchy 
associated with the finite-gap solutions to KP or KdV hierarchy. 

A single puncture is at $P_{0}$, whose local coordinate is, say,  $\xi(P_{0})=\xi_{0}=0$.
One can introduce meromorphic $1$-differentials with the poles at $\xi_{0}=0$ such that
\begin{equation}
d\Omega_{n}= \left( \xi^{-n-1}+ O(1)\right) d\xi,\ (P\rightarrow P_{0})\ \ n\ge 1
\label{dOmega}
\end{equation}
This condition determines $d\Omega_{n}$ up to arbitrary linear combination of $g$ holomorphic
differentials $dw_{i}\ (i=1,\cdots g)$. There are two ways to fix this ambiguity.

The first way is to impose that $d\Omega_{n}$ has vanishing $A$-periods,
\begin{equation}
\oint_{A_{i}}d\Omega_{n}=0,\ \ \forall i,n.
\end{equation}

We choose $P', P$ near $P_{0}$, and express local coordinates by $\zeta \equiv\xi(P'),\ \xi\equiv\xi(P)$.
It is known from the theory of Riemann surfaces that the generating functional is given by
\begin{equation}
W(\xi,\zeta)=\sum_{n=1}^{\infty}n\zeta^{n-1}d\zeta d\Omega_{n}(\xi)+ \cdots,
\end{equation}
This functional is explicitly given by 
\begin{equation}
W(\xi,\zeta)=\partial_{\xi}\partial_{\zeta}E(\xi,\zeta)
\end{equation}
where $\displaystyle E(\xi,\zeta)=\frac{\theta_{\ast}(\vec{\xi}-\vec{\zeta})}{\nu_{\ast}(\xi)\nu_{\ast}(\zeta)}$ (for genus $g=1$) is the Prime form~\cite{Mir4,Mir5}.

If we take the limit of $P\rightarrow P'\ (\xi\rightarrow\zeta)$, then 
\begin{equation}
W(\zeta, \xi)\sim \frac{d\zeta d\xi}{(\xi-\zeta)^{2}}+O(1)=\sum_{n=1}^{\infty}n\frac{d\xi}{\xi^{n+1}}\zeta^{n-1}d\zeta +O(1)
\end{equation}
and we have the second order pole near $P$.

The second way   is to impose the condition
\begin{equation}
\frac{\partial d\hat{\Omega}_{n}}{\partial\ \mathrm{moduli}}=\mathrm{holomorphic}.
\label{holocon}
\end{equation}
We restrict ourselves to the case where the number of moduli is equal to the genus of the spectral curve. This case is the one
in the SW theory. Following~\cite{Don1}-~\cite{War1},~\cite{Taka2},\cite{Taka1}-\cite{Mor3}, we introduce the generating functional for $d\hat{\Omega}$ with infinitely auxiliary parameters $T_{n}$,
\begin{equation}
dS=\sum_{n\ge 1}T_{n}d\hat{\Omega}_{n}=\sum_{i=1}^{g}\alpha^{i}dw_{i}+\sum_{n\ge 1}T_{n}d\Omega_{n}.
\label{predif}
\end{equation}
The periods
\begin{equation}
\alpha^{i}=\oint_{A_{i}}dS
\end{equation}
can be considered as the local coordinates of the moduli space. In particular,
\begin{equation}
\frac{\partial dS}{\partial \alpha^{i}}=dw_{i},\ \ \ \frac{\partial dS}{\partial T^{n}}=d\Omega_{n}
\end{equation}
and
\begin{equation}
T_{n}=\mathrm{res}_{\xi=0}\xi^{n}dS. 
\end{equation}

Equation (\ref{dOmega}) should be related with the potential differntail $dS$,
\begin{equation}
d\Omega_{n}=\frac{\partial dS}{\partial t_{n}}.
\end{equation}

Now we introduce the \textit{prepotential} $\mathcal{F}(\alpha^{i}, T^{n})$ such that
\begin{equation}
\frac{\partial \mathcal{F}}{\partial \alpha^{i}}=\oint_{B^{i}}dS,\ \ \
\frac{\partial\mathcal{F}}{\partial T_{n}}=\frac{1}{2\pi in}\mathrm{res}_{0}\xi^{-n}dS
\end{equation}
The $\tau$-function of the Whitham hierarchy is defined as
\begin{equation}
\log \mathcal{T}_{\mathrm{Whitham}}=\int_{\Sigma}d\bar{S}\wedge dS
\end{equation}
and the prepotential is given by~\cite{Kri1}
\begin{equation}
\mathcal{F}=\log \mathcal{T}_{\mathrm{Whitham}}.\label{pre-tau}
\end{equation}
 
\subsection{Whitham Hierarchy and Toda-chain Hierarchy} 

The Toda chain hierarchy is different from the KP/KdV hierarchy as considered above: 
The spectral curve for the Toda-chain is given by $\displaystyle w+\frac{\Lambda^{2N}}{w}=P_{N}(\lambda)$. This curve
has two punctures $w=0,\infty$.

For the spectral curve (\ref{cc1}), there exists a function $w$ with the $N$th-order pole and zero at two points 
$\lambda=\infty_{\pm}$, where $\pm$ labels two sheets of the curve in Eq(\ref{cc1}), 
$w(\lambda=\infty_{+})=\infty,\ w(\lambda=\infty_{-})=0$. Corresponding to the two punctures, there are two
families of the differentials $d\Omega_{n}$,: $d\Omega_{n}^{+}$ with poles at $\infty_{+}$ and $d\Omega_{n}^{-}$
with poles at $\infty_{-}$. Note that there are no differentials $d\hat{\Omega}^{\pm}$ because the condition (\ref{holocon}) requires
$d\hat{\Omega}_{n}$ to have poles at both punctures.

The differentials $d\hat{\Omega}_{n}$ have the form~\cite{Marshakov:1996ae}
\begin{equation}
d\hat{\Omega}_{n}=R_{n}(\lambda)\frac{dw}{w}=P_{+}^{n/N}(\lambda)\frac{dw}{w}
\end{equation}
where the polynomial $R_{n}(\lambda)$ is of the degree $n$ in $\lambda$ and 
$(\sum_{k=-\infty}^{\infty}a_{k}\lambda^{k})_{+}=\sum_{k=0}^{\infty}a_{k}\lambda^{k}$. Then, the formalism of the previous subsection
is applicable for $\xi=w^{\mp 1/N}$. 

\subsection{Seiberg-Witten Theory and Whitham Hierarchy}
We include additional equations involving $dw_{i},\ \alpha_{i}$ to the Whitham equations (\ref{WhithamEq}):
\begin{equation}
\frac{\partial d\Omega_{n}}{\partial \alpha_{i}}=\frac{\partial dw_{i}}{\partial t_{n}},\ \ \ 
\frac{\partial dw_{i}}{\partial \alpha_{j}}=\frac{\partial dw_{j}}{\partial \alpha_{i}}.
\end{equation}
These equations are solved by using the differential $dS$, 
\begin{equation}
\frac{\partial dS}{\partial \alpha_{i}}=dw_{i}.
\end{equation}

As is Eq.(\ref{predif}), generating functionals $d\hat{\Omega}_{n}$ are described with auxiliary parameters $T_{n}$,
\begin{equation}
dS=\sum_{n\ge 1}T_{n}d\hat{\Omega}_{n}=\sum_{i=1}^{g}\alpha^{i}dw_{i}+\sum_{n\ge 1}t^{n}d\Omega_{n}
\end{equation}
The periods are given by
\begin{equation}
\alpha^{i}=\oint_{A_{i}}dS
\end{equation}
In particular, we have
\begin{equation}
\frac{\partial dS}{\partial \alpha^{i}}=dw_{i},\ \ \ \frac{\partial dS}{\partial t^{n}}=d\Omega_{n}.
\end{equation}
Now we introduce the \textit{prepotential} $\mathcal{F}(\alpha^{i}, T^{n})$ such that
\begin{equation}
\frac{\partial \mathcal{F}}{\partial \alpha^{i}}=\oint_{B^{i}}dS,\ \ \
\frac{\partial\mathcal{F}}{\partial T_{n}}=\frac{1}{2\pi in}\mathrm{res}_{0}\xi^{-n}dS
\end{equation}
and the Seiberg-Witten differential $dS_{SW}=d\hat{\Omega}_{2}$
\begin{equation}
dS|_{t_{n}=\delta_{n,1}}=dS_{SW},\ \ \ \alpha^{i}|_{t_{n}=\delta_{n,1}}=a^{i},\ \ \ \alpha_{i}^{D}|_{t_{n}=\delta_{n,1}}=a_{i}^{D}.
\end{equation}

\section{Noncommutative KP Hierarchy}
The derivation of noncommutative KP hierarchy is almost parallel to the commutative case.   
We follow the formulation of the $\bar{\partial}$-problem~\cite{WW}. 

Let us consider an infinite noncommutative space-time (the symbol $\wedge$  on the letter means an operator),
\begin{eqnarray}
&&[\hat{t}_{1},\hat{t}_{2}]=i\theta_{1},\ \ [\hat{t}_{3},\hat{t}_{4}]=i\theta_{2},\cdots,[\hat{t}_{2n-1},\hat{t}_{2}]=i\theta_{n},
n=1,2,\cdots,\infty  \\
&&\left[\hat{x}_{2n-1},\hat{x}_{2m}\right]=0,\ [\hat{x}_{2n-1},\hat{x}_{2m-1}]=0,\ \ \mathrm{for}\ n\neq m
\end{eqnarray}
with the Hermitian condition, $\hat{x}_{k}^{\dagger}=\hat{x}_{k}$ and $\theta_{n}=0, n\gg 0$ for simplicity.

The analytic-bilinear approach begins with a nonlocal $\bar{\partial}$-problem,
\begin{equation}
\frac{\partial}{\partial\bar{\alpha}}\hat{\chi}(\alpha,\beta)=2\pi i\delta(\alpha-\beta)
+\int\!\int_{C}d\alpha''\wedge d\bar{\alpha}''\hat{\chi}(\alpha'',\beta)\hat{R}(\alpha'',\bar{\alpha''};\alpha',\bar{\alpha'}),
\end{equation}
where $\alpha,\beta\in\mathbf{C}$ are spectral parameters, the bar indicates the complex conjugate and $\hat{R}$ is called 
the spectral transform operator.

NC generalized analytic-bilinear identity holds
\begin{equation}
\oint_{\partial G}d\beta^{'}\hat{\chi}(\beta^{'},\beta;\hat{g}_{1})\hat{g}_{1}(\beta^{'})\hat{g}_{2}(\beta^{'})^{-1}
\hat{\chi}(\alpha,\beta^{'};\hat{g}_{2})=0. \label{ABI}
\end{equation}
where $G$ is a domain in the complex plane,and $\hat{g}(\beta)=g(\beta,\hat{t})$.

\subsection{NC KP Equation}
We consider the relation,
\begin{equation}
\hat{g}_{1}(\beta^{'})\hat{g}_{2}(\beta^{'})=\exp \left(\sum_{k\ge 1}(\hat{t}_{k}-\hat{t^{'}}_{k})(\beta^{'})^{-k}\right)
=\exp\left(\sum_{k\ge 1}\frac{1}{k}(a^{k}-b^{k})(\beta^{'})^{-k}\right)
\end{equation}
where $a,b\in G$ are usual c-numbers. Note that the differences of operator coordinates are replaced by c-numbers. By 
performing the residue integral of (\ref{ABI}), we obtain
\begin{eqnarray}
\frac{\beta-a}{\beta-b}\chi(\alpha,\beta;\hat{t}+[a])-\frac{\alpha-a}{\alpha-b}\chi(\alpha,\beta;\hat{t}+[b]) \nonumber \\ 
=(a-b)\chi(b,\beta;\hat{t}+[b])\chi(\alpha,b;\hat{t}+[a])  \label{RF}
\end{eqnarray}
where $[a]\equiv (a,\frac{a^{2}}{2},\cdots, \frac{a^{k}}{k},\cdots)$ and $\hat{t}\equiv (\hat{t}_{1},\hat{t}_{2},\cdots)$.

Introducing the Cauchy-Baker-Akhiezer (CBA) function,
\begin{equation}
\psi(\alpha,\beta;\hat{t})\equiv \hat{g}(\beta)^{-1}\chi(\alpha,\beta;\hat{t})g(\alpha,\hat{t})\label{CBA}
\end{equation}
and substituting (\ref{CBA}) into (\ref{RF}), we find
\begin{equation}
\psi(\alpha,\beta;\hat{t}+[a])-\psi(\alpha,\beta;\hat{t})=a\psi(0,\beta;\hat{t})\psi(\alpha,0;\hat{t}+[a])
\label{NCKP}
\end{equation}
Equation (\ref{NCKP}) contains the whole NC KP and NC mKP (modified KP) hierarchies. We can obtain the NC KP hierarchy by expanding 
Eq. (\ref{NCKP}) in the power of the c-number parameter $a$. Another way to derive NC KP hierarchy is to construct the NC Sato theory. The derivation of the NC KP hierarchy through NC Sato theory is discussed in detail in~\cite{WW}. In what follows, we shall restrict 
to the NC KP equation and derive the triple-product relation (WDVV equations like) which 
the $\tau$-function (operator valued) representation of NC KP hierarchy satisfies.

For example, the Lax-pair for the NC KP equation~\cite{Paniak:2001fc} is given by
\begin{eqnarray}
\hat{\partial}_{y}\hat{f}-\hat{\partial}_{xx}\hat{f}&=&\hat{u}\hat{f} \nonumber \\
\hat{f}_{t}-\hat{\partial}_{xxx}\hat{f}&=&\frac{3}{2}\hat{u}\hat{\partial}_{x}\hat{f}+\frac{3}{4}(\hat{u}_{x}
+\hat{\partial}_{x}^{-1}\hat{\partial}_{y})\hat{f}\label{NCKP2}
\end{eqnarray}
where $\hat{u}\equiv -2\psi(0,0,\hat{x})_{x},\ \hat{f}\equiv\int\psi(\alpha,0;\hat{x})\rho(\alpha)d\alpha$ with an arbitrary 
function $\rho(\alpha)$. One obtains the NC KP equation from the compatibility of Eq. (\ref{NCKP2}),
\begin{equation}
\hat{u}_{t}-\frac{3}{4}\hat{\partial}_{x}^{-1}\hat{u}_{yy}-\frac{1}{4}\hat{u}_{xxx}-\frac{3}{4}(\hat{u}_{x}\hat{u}+\hat{u}\hat{u}_{x})
+[\hat{u},\hat{\partial}_{x}^{-1}\hat{u}_{y}]=0
\end{equation}
  
We will construct $\tau$-function for NC KP hierarchy. First, Let us introduce a generalized shift operator 
$T_{\alpha}f(\hat{t})\equiv f(\hat{t}-[\alpha])$. The operator valued function, $\chi(\alpha,\beta;\hat{t})$, has a 
$\tau$-function representation,
\begin{equation}
\chi(\alpha,\beta;\hat{t}+[\alpha])=\frac{1}{\alpha-\beta}\tau(\hat{t}+[\beta]-[\alpha])\tau^{-1}(\hat{t})
\end{equation}
with 
\begin{equation}
T_{\beta}\tau(\hat{t})\frac{1}{T_{\alpha}\tau(\hat{t})}T_{\gamma}\tau(\hat{t})
=T_{\gamma}\tau(\hat{t})\frac{1}{T_{\alpha}\tau(\hat{t})}T_{\beta}\tau(\hat{t}).
\label{(5.2)}
\end{equation}
This functional relation is called (Hirota) triple-product relation~\cite{WW}

The proof of the relation (\ref{(5.2)}) is as follows. Choosing $\beta=a,\ \alpha=a$ in Eq.(\ref{RF}) and changing the variables from $\hat{t}$ to $\hat{t}-[a]$ and $\hat{t}-[b]$, one obtains
\begin{eqnarray}
    \{(\beta-a)\chi(\beta,a;\hat{t}-[a]) \}\{(\alpha-\beta)\chi(\alpha,\beta;\hat{t}-[\beta])\}=(\alpha-a)\chi(\alpha,a;\hat{t}-[a]) 
   \label{(5.5)} \\
    \{(\alpha-\beta)\chi(\alpha,\beta;\hat{t}+[\alpha]) \}\{(a-\alpha)\chi(a,\alpha;\hat{t}+[a])\}=(a-\beta)\chi(a,\beta;\hat{t}+[a]).
   \label{(5.6)} 
\end{eqnarray}
Set
\begin{equation}
\chi(\alpha,\beta;\hat{t}+[\alpha])=\frac{1}{\alpha-\beta}\tau(\hat{t}+[\beta]-[\alpha])\frac{1}{\tau(\hat{t})}
\label{(5.7)}
\end{equation}
where $\hat{\tau}$ is an arbitrary (operator-valued) function. One can easily check that Eq.(\ref{(5.6)}) is automatically 
satisfied by Eq.(\ref{(5.7)}). Substitution of  Eq.(\ref{(5.7)}) into Eq.(\ref{(5.5)}) yields Eq.(\ref{(5.2)}). 

\subsection{Triple-product Difference Equation}
We will see Eq.(\ref{(5.2)}) is cast into the same form as WDVV equations (\ref{WDVV}). We define 
$\tau(\hat{t}+m[\alpha]+k[\beta]+n[\gamma])\equiv \hat{\tau}(m,k,n)$, and denote 
$\hat{F}_{k}=\hat{\tau}(m,k-1,n),\ \hat{F}_{m}=\hat{\tau}(m-1,k,n),\ \hat{F}_{n}=\hat{\tau}(m,k,n-1)$.
Then, Eq.(\ref{(5.2)}) yeilds the following relation,
\begin{equation}
\hat{F}_{m}\frac{1}{\hat{F}_{k}}\hat{F}_{n}=\hat{F}_{n}\frac{1}{\hat{F}_{k}}\hat{F}_{m} \label{(6.8)}.
\end{equation}
Note that the relation(\ref{(6.8)}) arises from the noncommutative property and reduces to a trivial one 
in the commutative limit. 

\subsection{Deformation}
In this subsection, we construct the triple-product relation from the commutative case. A key idea is that $N\times N$ 
matrices are identified as an operator in the large $N$ limit.
 
The WDVV equations (\ref{WDVV}) are described via the third derivative of the prepotential. Substituting  
Eq. (\ref{pre-tau}) to Eq. (\ref{WDVV}), we have an explicit expression, 
\begin{equation}
\mathcal{P}\tau_{i}\frac{1}{\mathcal{P}\tau_{k}}\mathcal{P}\tau_{j}
=\mathcal{P}\tau_{j}\frac{1}{\mathcal{P}\tau_{i}}\mathcal{P}\tau_{j},
\label{tauWDVV}
\end{equation}
where 
\begin{eqnarray}
&&\partial_{i}=\frac{\partial}{\partial a_{i}},\ \ \tau_{i}=\frac{\partial \tau}{\partial a_{i}}, 
\ \ \tau_{ij}=\partial_{i}\partial_{j}\tau, \\
&&(\mathcal{P})_{jk}=2\frac{\tau_{j}\tau_{k}}{\tau^{3}}-\frac{\tau_{jk}}{\tau^{2}}
-\frac{\tau_{j}}{\tau^{2}}\partial_{k}
-\frac{\tau_{k}}{\tau^{2}}\partial_{j}+\frac{1}{\tau}\partial_{j}\partial_{k}.
\end{eqnarray}
We introduce the cut-off parameter $\alpha, \beta, \gamma\in \mathbf{C}$ to identify the derivative of the $\tau$-function
as a difference of two points. The derivative are replaced by the following form  
\begin{equation}
\frac{\partial \tau}{\partial a_{i}}\rightarrow \frac{\tau(a_{i}+m\alpha)-\tau(a_{i}+(m-1)\alpha)}{\alpha}.
\end{equation}

We consider the case where the number of the moduli parameter is infinite. The matrix whose elements are the combination
of functions and derivatives may be considered as an operator $\hat{\mathcal{P}}$. In this procedure, Eq. (\ref{tauWDVV}) has the 
form
\begin{equation}
\hat{\mathcal{P}}\kappa(m-1,k,n)\frac{1}{\hat{\mathcal{P}}\kappa(m,k-1,n)}\hat{\mathcal{P}}\kappa(m,k,n-1)
=\hat{\mathcal{P}}\kappa(m,k,n-1)\frac{1}{\hat{\mathcal{P}}\kappa(m,k-1,n)}\hat{\mathcal{P}}\kappa(m-1,k,n)
\end{equation}
where
\begin{eqnarray}
\kappa(m-1,k,n)=\tau(a_{I}+m\alpha,a_{K}+k\beta,a_{J}+n\gamma)-\tau(a_{I}+(m-1)\alpha,a_{K}+k\beta,a_{J}+n\gamma), \\
\kappa(m,k-1,n)=\tau(a_{I}+m\alpha,a_{K}+k\beta,a_{J}+n\gamma)-\tau(a_{I}+m\alpha,a_{K}+(k-1)\beta,a_{J}+n\gamma), \\ 
\kappa(m,k,n-1)=\tau(a_{I}+m\alpha,a_{K}+k\beta,a_{J}+n\gamma)-\tau(a_{I}+m\alpha,a_{K}+k\beta,a_{J}+(n-1)\gamma).
\end{eqnarray}
To deform the commutative $\tau$-function to the noncommutative $\tau$-function, we assume that the operator $\hat{\mathcal{P}}$ 
changes the commutative function to the noncommutative one. 

If we set $\hat{\kappa}(m,k,n)\equiv\hat{\mathcal{P}}\kappa(m,k,n)$, we find the triple-product relation
\begin{equation}
\hat{\kappa}(m-1,k,n)\frac{1}{\hat{\kappa}(m,k-1,n)}\hat{\kappa}(m,k,n)=\hat{\kappa}(m,k,n-1)\frac{1}{\hat{\kappa}(m,k-1,n)}\hat{\kappa}(m-1,k,n)
\end{equation}

In the above procedure, we shift the moduli parameter by the c-number $\alpha,\beta,\gamma$. In the noncommutative triangle-product 
relation, the ``time " parameter (operator-valued) is shifted by the c-number. Through the Whitham formulation, the moduli $\alpha_{i}$
is not independent of the time parameter $t_{m}$ and vice versa. Therefore, in the deformation of the commutative function to the 
noncommutative one via the operator $\hat{\mathcal{P}}$, the shift in the moduli parameter is translated to the c-number shift in the 
operator-valued time parameter. 

\section{Concluding Remarks}
It is interesting to compare to Eq.(\ref{WDVV}) and Eq.(\ref{(6.8)}). First, matrices for the WDVV equations are constructed 
from the derivatives of the prepotential in the continuous variables (quantum moduli), while functions in Eq.(\ref{(6.8)}) are 
expressed in terms of discrete variables. Second, the matrix representation of the $\tau$-function of the NC KP hierarchy has an 
infinite dimension. When we consider the situation where the difference of two noncommutative coordinates is a c-number, 
Eq.(\ref{(6.8)}) means that there exists an independent ``direction" on the noncommutative space. Third, the $\tau$-function for the NC KP hierarchy is an operator-valued function.

To examine an operator-valued $\tau$-function for the noncommutative Toda hierarchy or noncommutative Whitham hierarchy may be useful to see the correspondence of the spectral curve for the Seiberg-Witten theory. Another way to reveal the relation between the WDVV equations and the noncommutative theory may be in the string-noncommutative context. These subjects will be discussed elsewhere. 

\section*{Acknowledgments}

The authors would like to thank J.~Ieda and Y.~Uchiyama for useful comments.

\end{document}